\documentclass{iau}

\usepackage{amsmath}
\usepackage{graphicx}
\usepackage{multirow}
\usepackage{hyperref}

\renewcommand{\orcid}[1]{\,\href{https://orcid.org/#1}{\includegraphics[height=10pt]{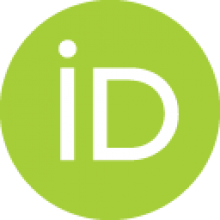}}}

\begin{document}

\lefttitle{Thibaut L. François}
\righttitle{Jacobi capture as a pathway to off-center massive black hole binaries in dwarf galaxies}

\volno{398}
\jnlPage{1}{7}
\jnlDoiYr{2025}
\doival{10.1017/xxxxx}

\aopheadtitle{Proceedings IAU Symposium}
\editors{Hyung Mok Lee, Rainer Spurzem \& Jongsuk Hong, eds.}

\title{Jacobi capture as a pathway to off-center massive black hole binaries in dwarf galaxies}

\author{Thibaut L. François\orcid{0009-0001-0314-7038}}
\affiliation{School of Mathematics and Physics, University of Surrey, Guildford, GU2 7XH, UK}

\begin{abstract}
The recent detection of high-redshift supermassive black holes with JWST has renewed interest in the processes driving black hole growth. At the same time, both simulations and observations point to a widespread population of off-center intermediate-mass black holes in dwarf galaxies. Their ability to merge outside galactic centers may play a key role in shaping black hole mass assembly. Here, we investigate the dynamics of off-center black holes in dwarf galaxies hosting cored dark matter haloes, where long dynamical friction timescales and core stalling naturally arise. By embedding off-center black holes into an idealized galactic potential and scanning a wide range of orbital configurations, we assess the likelihood of close interactions through Jacobi capture. We find that captures occur in about $13\%$ of cases. Such captures, possibly sustained within compact stellar systems like stripped nuclei or globular clusters, represent a crucial first step toward assembling massive black hole binaries beyond galactic centers.
\end{abstract}

\begin{keywords}
chaos – gravitation – methods: numerical – galaxies: dwarf – galaxies: kinematics and dynamics
\end{keywords}

\maketitle

\section{Introduction}

\begin{figure}[t]
  \centering
  \includegraphics[scale=.25]{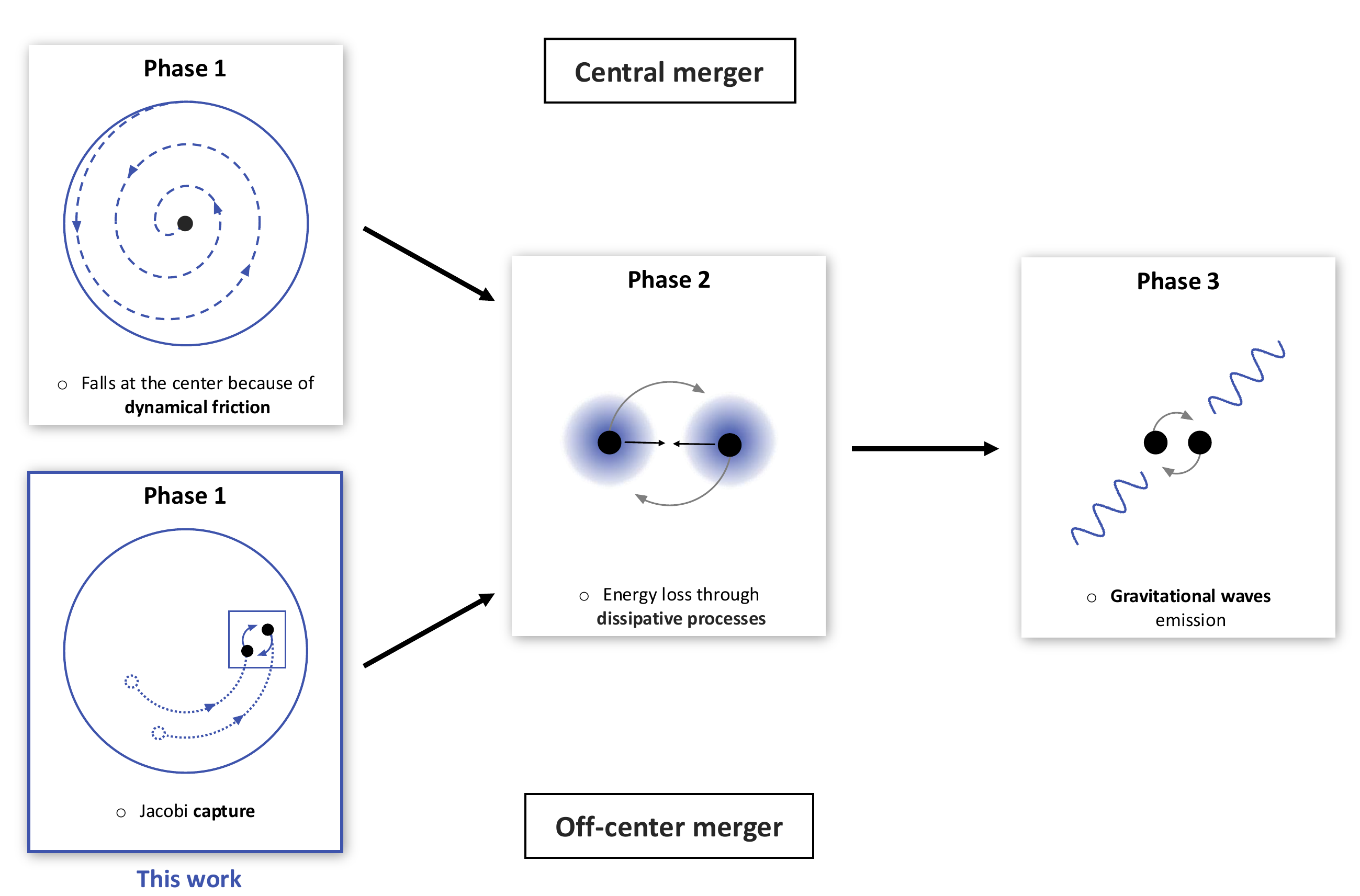}
  \caption{Top: standard scenario of a black hole merger occurring at the galactic center. Bottom: new scenario of an off-center merger.}
  \label{schema_fusion}
\end{figure}

In recent decades, observations have revealed massive black holes not only in large galaxies but also in dwarf systems, hinting at a common evolutionary link between black holes and their host galaxies \citep[e.g.,][]{Gebhardt2000,Reines2013}. Traditionally, black hole mergers are thought to proceed through a three-stage sequence \citep{Begelman1980}, beginning with the inspiral toward the galactic center driven by dynamical friction (top panel of Fig.~\ref{schema_fusion}). However, in low-density environments such as dwarf galaxies, this process can be severely delayed, with inspiral times exceeding a Hubble time, or even stalled due to core stalling \citep[][]{Kaur2018, Banik2021}. As a result, a significant population of intermediate-mass black holes may remain off-center \citep[][]{Bellovary2019, Bellovary2021, Pfister2019, Reines2020}. Understanding whether these off-center black holes can still merge is therefore critical for models of black hole mass growth. Here, we examine the initial stage of such mergers: Jacobi capture, a variation of the three body problem \citep{PetitHenon86} which occurs when two black holes enter each other's sphere of influence (lower panel of Fig.~\ref{schema_fusion}; see \citealt{Francois2024} for details). We simulate pairs of off-center black holes in a simplified dwarf galaxy potential, providing general insights that can guide more detailed future studies.

\section{Method}

\begin{figure}[t]
  \centering
  \includegraphics[scale=.4,trim=3cm 4cm 3cm 3cm, clip]{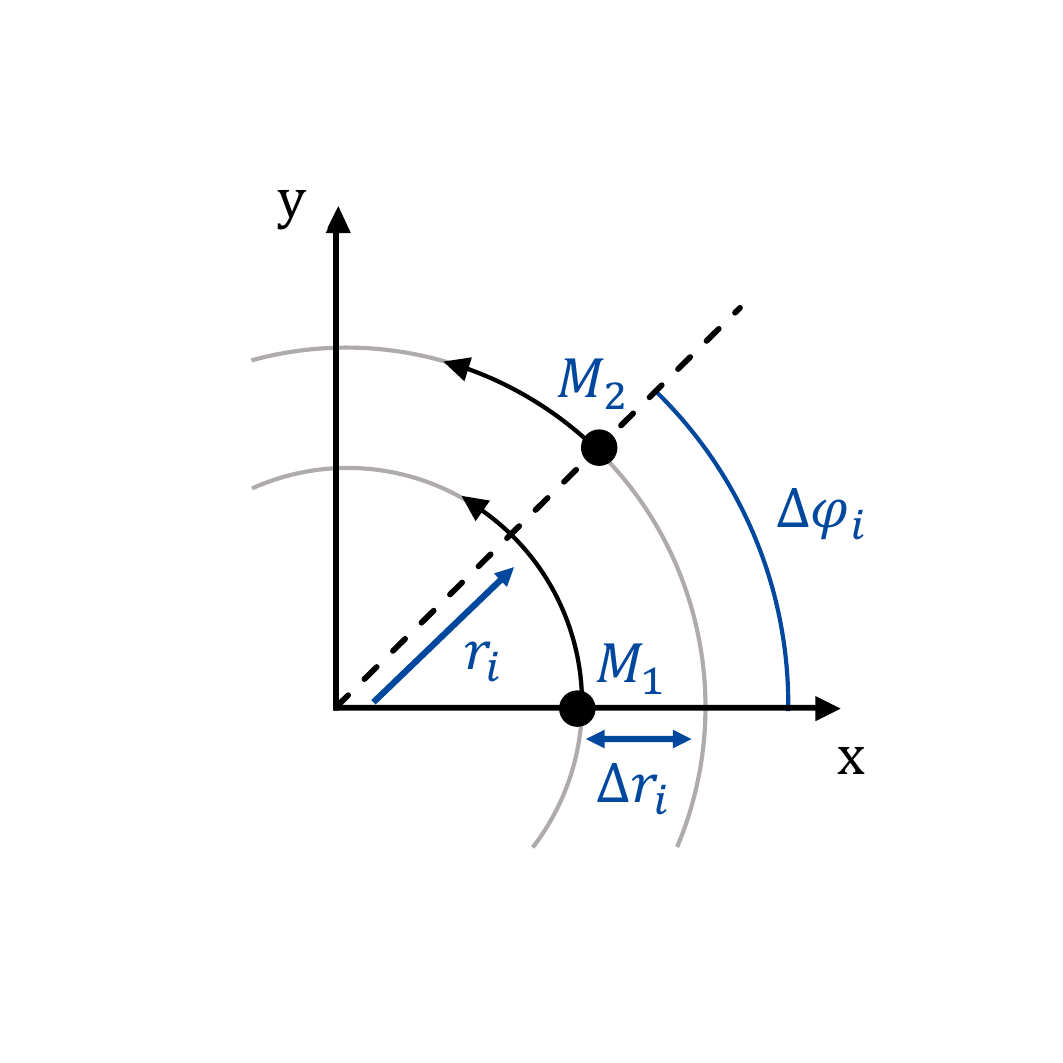}
  \caption{Parameters defining the initial configuration: $r_i$ is the initial radius of the inner black hole, $\Delta r_i$ the radial separation between the two black holes, $\Delta \varphi_i$ their phase offset, and $M_1, M_2$ their respective masses.
}
  \label{Para}
\end{figure}

We adopt a simplified setup in which two black holes are integrated on coplanar circular orbits within an external gravitational potential modeling the cored dark matter halo of a dwarf galaxy. The five-dimensional parameter space of the encounter (cf. Fig.~\ref{Para}) is systematically explored to evaluate the likelihood of Jacobi captures and to assess the influence of each parameter on the capture process.
In a second step, we reduce the number of free parameters to estimate the capture probability more realistically. To do so, the black holes are initially placed at their core-stalling radii, and their masses are drawn from a cosmologically motivated merger sequence using the semi-analytical code \textsc{SatGen} \citep[][]{satgen}. To classify an interaction as a capture, two criteria must be satisfied: (i) the binding energy of the pair must be negative; and (ii) the separation between the black holes must fall below the binary's sphere of influence, estimated from the Lagrange points. In cored density profiles, these points undergo a bifurcation that modifies the orbital dynamics, leading to an influence radius typically about twice as large as predicted by the distant-tide approximation in the core region.

\section{Results}

\begin{figure}[t]
  \centering
  \includegraphics[scale=.2,trim=0 0 0 30cm, clip]{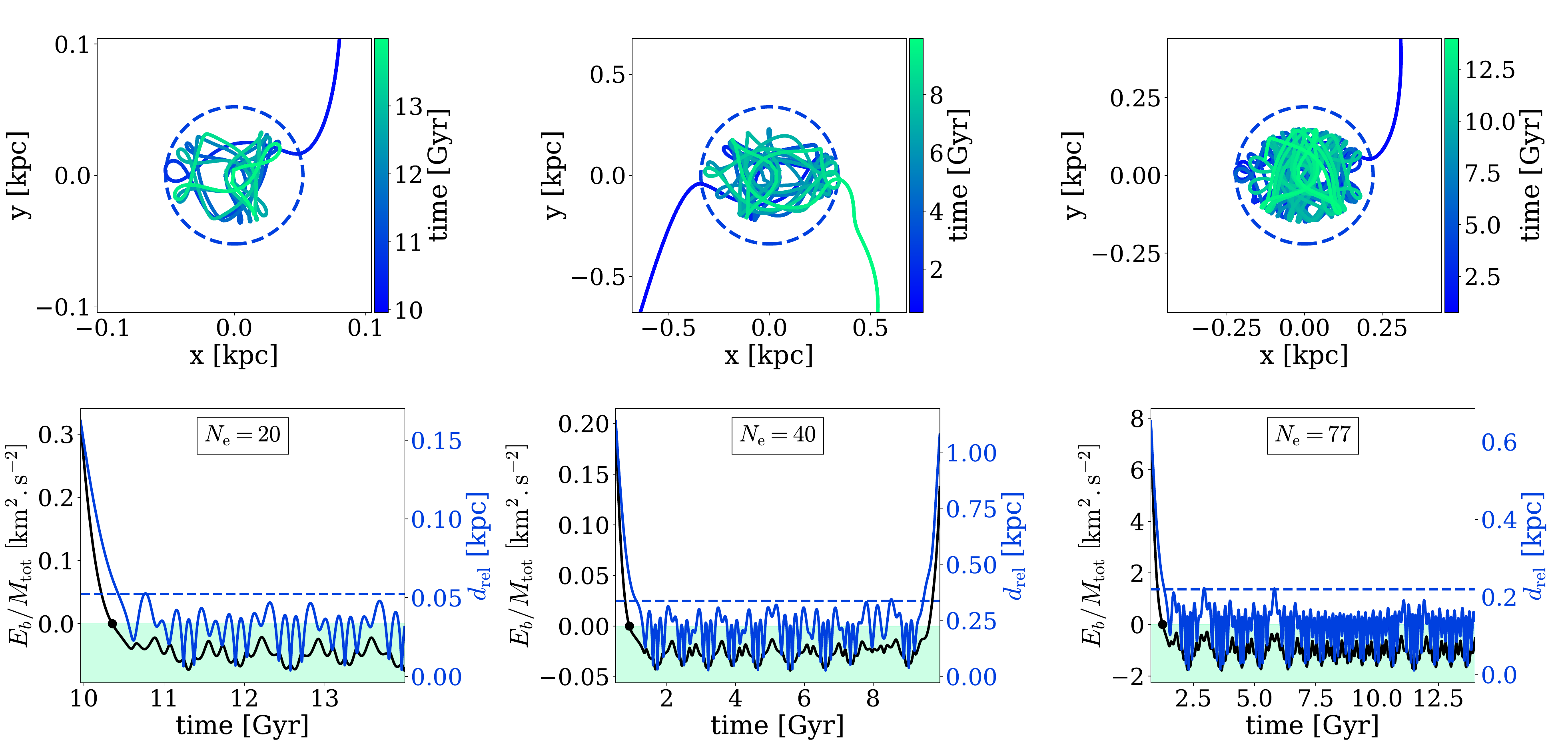}
  \caption{Examples of Jacobi captures with different numbers of close encounters ($N_\mathrm{e}$) between black holes. Top row: relative coordinates between black holes during the capture. The reference frame is centered on the most massive black hole and rotates with it. The binary influence radius is marked by the dotted circle. Bottom row: the black line represents the binding energy ($E_\mathrm{b}$) per unit mass of the binary ($M_\mathrm{tot}$), with negative values shown in the colored area. The blue curve represents the relative distance ($d_\mathrm{rel}$) between the black holes. The binary influence radius is indicated by a dashed blue horizontal line.}
  \label{trajRH}
\end{figure}

\begin{figure}[t]
  \centering
  \includegraphics[scale=.41]{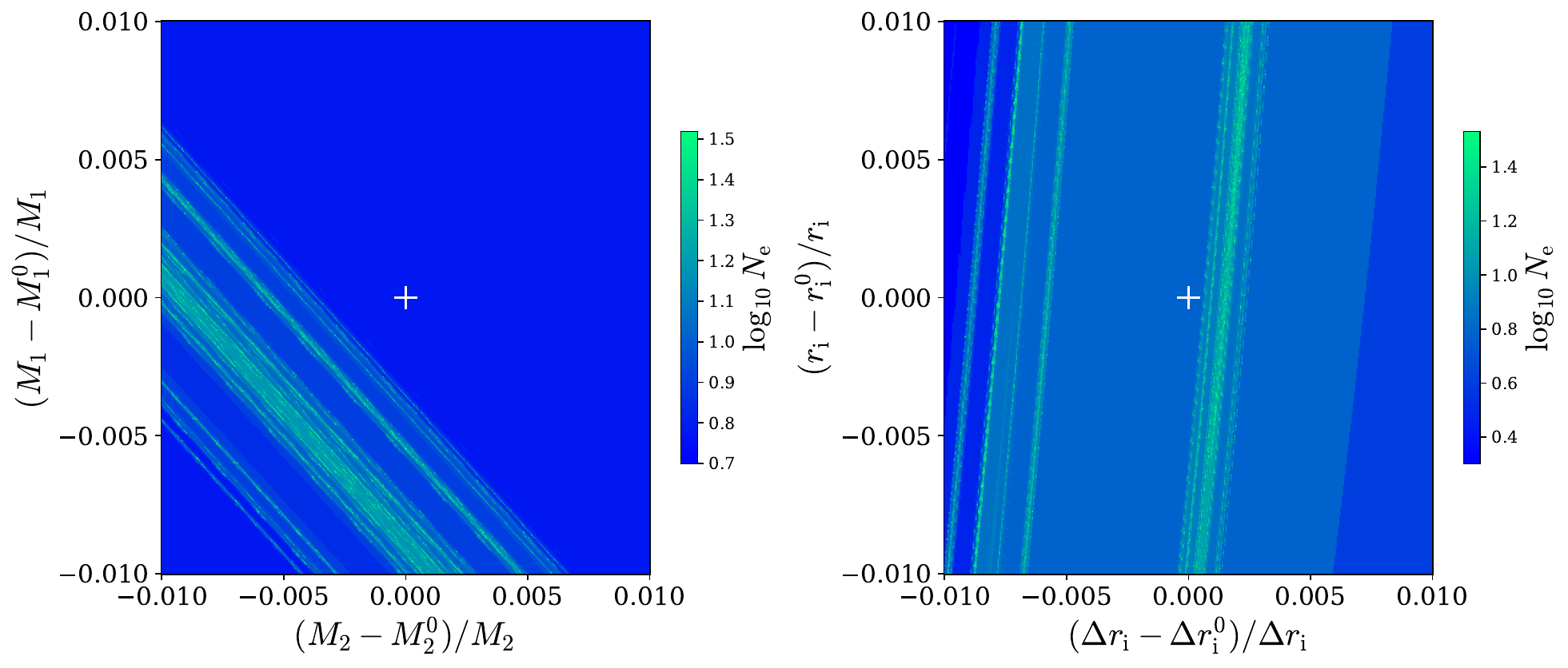}
  \caption{Influence of the initial conditions on the number of close encounters ($N_\mathrm{e}$). We choose a reference capture involving five close encounters between the black holes and look at small variations of initial conditions.
  Central panel: variations of $N_\mathrm{e}$ as a function of the initial masses ($M_1, M_2$). The explored values of ($M_1, M_2$) differ by $\pm1\%$ from the reference configuration, ($M_1^0, M_2^0$).
  Right panel: variations of $N_\mathrm{e}$ with respect to the initial orbital parameters ($r_\mathrm{i}, \Delta r_\mathrm{i}$). Here, ($r_\mathrm{i}, \Delta r_\mathrm{i}$) are also varied within $\pm1\%$ of the reference values ($r_\mathrm{i}^0, \Delta r_\mathrm{i}^0$).
  In both panels, the reference case corresponds to the point ($0,0$). The strong variations of $N_\mathrm{e}$ with initial conditions is suggestive of chaotic motion.}
  \label{droite}
\end{figure}

Figure~\ref{trajRH} presents three examples of Jacobi captures. In each case, the black holes enter the binary's sphere of influence and undergo a sequence of close gravitational encounters before eventually separating again. These captures are temporary, although in some cases (e.g. left and right panel of Fig.~\ref{trajRH}) the interaction is still ongoing when the simulation stops at 14 Gyr, so that the unbinding, while expected, is not observed. Figure~\ref{droite} illustrates the sensitivity of the number of close encounters to small variations in the initial conditions. Here, a reference capture is changed by less than $\pm 1\%$ in either the black hole masses ($M_1, M_2$) or the initial radii ($r_\mathrm{i}, \Delta r_\mathrm{i}$). The outcome varies drastically, with the number of close encounters changing significantly even under tiny variations. This strong dependence with initial conditions is suggestive of chaotic motion. We also examined the eccentricity distribution of the captures and found that they are typically highly eccentric, with nearly all events having $e>0.5$ and a large fraction exceeding $e>0.9$. Regarding trends with initial parameters, we find that captures are favored when the outer black hole is more massive than the inner one. In this configuration, the lighter black hole approaches the heavier one at a larger radius, where the potential well is shallower and orbital velocities are lower. As a result, their relative velocity at encounter is reduced, which facilitates the capture process. 
In addition, captures tend to cluster in bands of radial separation $\Delta r_\mathrm{i}$. At larger separations, the black holes remain too distant to interact, while at smaller separations their orbital frequencies are nearly identical, preventing sufficient phase drift to enable capture within a Hubble time. Finally, by reducing the number of free parameters—placing the black holes at their core-stalling radii and drawing their masses from cosmological merger trees generated with \textsc{SatGen}—we obtain an overall capture probability of about $13\%$.

\section{Conclusions}

In this study, we examined whether black holes in dwarf galaxies can merge outside galactic centers. Using a simplified setup, we tracked the formation of off-centered binaries through Jacobi captures and analyzed how this process depends on the initial conditions. Our results show that the stability and subsequent evolution of captures are highly sensitive to the initial conditions, with a capture being initiated in about $13\%$ of cases. While these events are temporary in the absence of dissipation, they could be stabilized by additional processes such as interactions within dense stellar systems. This suggests that stripped nuclei and globular clusters may be promising sites where captures contribute to black hole growth. More broadly, our findings indicate that such interactions could represent a non-negligible channel for the assembly and evolution of black holes across cosmic time.

\vspace{0.5cm}

\bibliographystyle{aasjournalv7}
\bibliography{Sample}

\end{document}